\documentclass[12pt]{article} 
\pdfoutput=1
\usepackage{epsfig}
\usepackage{graphicx}
\usepackage{jheppub}
\usepackage{amsmath}
\usepackage{amsfonts}
\usepackage{amssymb}
\usepackage{graphicx}
\usepackage{inputenc}
\usepackage{multirow}
\usepackage{color}
\usepackage{hyperref}
\usepackage[compat=1.0.1]{tikz-feynman}
\usepackage{subcaption}

\def\beq{\begin{equation}}
\def\eeq{\end{equation}}
\def\beqa{\begin{eqnarray}}
\def\eeqa{\end{eqnarray}}

\def\eq#1{Eq.~(\ref{#1})}

\newcommand{\fig}{Fig.~\ref}

\renewcommand{\epsilon}{\varepsilon}
\def \a{\alpha}

\def \d {{\rm d}}

\newcommand{\ord}{{\cal O}}
\newcommand{\RE}{{\rm Re}}

\def\ifm{\ifmmode}

\def \CG {{\cal G}}

\definecolor{darkgreen}{rgb}{0.0, 0.4, 0.13}

\newcommand\sss{\scriptscriptstyle}
\newcommand\as{\alpha_{\sss S}} 

\def \heft{{\rm\sss HT}}
\def \heftew{{\rm\sss HT-EW}}

\def \al #1 {\frac {\as({#1})}{\pi} }

\allowdisplaybreaks[1]

\preprint{MIT-CTP/5078}

\title{Mixed QCD-electroweak corrections to Higgs production via gluon fusion in
  the small mass approximation}

\author[a]{Charalampos Anastasiou,}
\author[a,1]{Vittorio Del Duca,}
\note{On leave from INFN, Laboratori Nazionali di Frascati, Italy.}
\author[a]{Elisabetta Furlan,}
\author[b]{Bernhard Mistlberger,}
\author[a]{Francesco Moriello,}
\author[a]{Armin Schweitzer,}
\author[a]{Caterina Specchia}
\affiliation[a]{Institute for Theoretical Physics, ETH Z\"urich, 8093 Z\"urich, Switzerland.}
\affiliation[b]{Center for Theoretical Physics, Massachusetts Institute of Technology, Cambridge, MA 02139, USA.}

\emailAdd{babis@phys.ethz.ch}
\emailAdd{delducav@itp.phys.ethz.ch}
\emailAdd{efurlan@phys.ethz.ch}
\emailAdd{bernhard.mistlberger@gmail.com}
\emailAdd{fmoriell@phys.ethz.ch}
\emailAdd{armin.schweitzer@phys.ethz.ch}
\emailAdd{specchic@phys.ethz.ch}

\abstract{
We compute the mixed QCD-electroweak corrections to the 
cross section for the production of a Higgs boson via gluon fusion, 
in the limit of a small mass of the electroweak gauge bosons.  
This limit is regular and we calculate it by setting the $W,Z$ masses to
zero in the Feynman rules for their propagators. 
Our analytic results provide an independent check, in a non-trivial
limit, of a recent exact computation for the three-loop 
mixed QCD and electroweak virtual corrections \cite{Bonetti:2017ovy}
and the corresponding contribution to the cross section in the
soft-virtual approximation~\cite{Bonetti:2018ukf}. 
From our calculation in the small mass approximation, we  can infer 
the second term in the expansion of the cross section around the
threshold limit with its exact dependence on the masses of the $W,\,Z$ bosons. 
Furthermore we find that in the small mass approximation the non-factorizable contributions from the real radiation, so far unknown for full gauge boson mass dependence, are modest in comparison to the known factorizable and virtual contributions to the full $\mathcal O (\alpha_s^3 \, \alpha^2)$ mixed QCD and electroweak cross-section.
This furnishes a new phenomenological test of  
estimates~\cite{Anastasiou:2008tj}  for the mixed QCD and electroweak
corrections, which were based on the hypothesis of factorization of
QCD and electroweak corrections.   
}

\keywords{Higgs physics, QCD, gluon fusion, electroweak corrections.}

\notoc
\begin{document}
\begin{flushright}
\vspace*{-25pt}
\end{flushright}
\maketitle
\allowdisplaybreaks 


\section{Introduction}
\label{sec:intro}

The discovery of the Higgs boson~\cite{Aad:2012tfa,Chatrchyan:2012xdj} at the Large Hadron Collider (LHC) at CERN
has been the ultimate success of the Standard Model (SM) of particle physics, and a terrific start for the LHC physics program.
With that, the LHC community has entered a phase of precision measurements with emphasis on the properties, couplings and
quantum numbers of the Higgs boson.
In parallel, the theory  community has made a decades long effort in the computation of the
Higgs production cross section via the dominant gluon-fusion production mechanism, at ever increasing accuracy.
The leading-order production cross section was computed in the 70's~\cite{Georgi:1977gs}, and
the next-to-leading-order (NLO) QCD corrections were computed in the 90's~\cite{Graudenz:1992pv,Spira:1995rr}. 
The NLO corrections are large ($\sim 80-100\%$) casting doubt whether 
a perturbative expansion in the strong coupling constant $\alpha_s$
would yield a reliable theoretical estimate of the cross section.

The next-to-next-to-leading-order (NNLO) QCD corrections~\cite{Harlander:2002wh,Anastasiou:2002yz,Ravindran:2003um}
were computed in the Higgs Effective Field Theory (HEFT), i.e. in the limit of a top quark
much heavier than the Higgs boson, while all other quarks are taken as massless~\cite{Chetyrkin:1997un,Chetyrkin:2005ia,Schroder:2005hy},
and turned out to be significant ($\sim 10-20\%$), but smaller
than the NLO corrections, indicating that the $\alpha_s$ expansion might be
stabilising. Recently, the next-to-next-to-next-to-leading-order (N$^3$LO) QCD corrections
in HEFT have been computed~\cite{Anastasiou:2015ema,Mistlberger:2018etf},
which turn out to be small ($\sim 4-6\%$)~\cite{Anastasiou:2016cez}, putting on 
solid ground QCD predictions in perturbation theory for the gluon fusion cross section.  
The N$^3$LO cross section shows a remarkable
stability with respect to the choice of renormalisation and
factorisation scale, with a typical scale variation of less than $\pm 2\%$. 
At this level of precision other effects are important, which are not
captured by the QCD perturbative expansion in the combined 
heavy top-quark/massless light-quarks limit of HEFT (see for example ref.~\cite{Anastasiou:2016cez} for a comprehensive study).  

Finite quark-mass effects for all flavours are known exactly 
through NLO~\cite{Spira:1995rr,Harlander:2005rq,Bonciani:2007ex,Anastasiou:2006hc,Anastasiou:2009kn}.
These amount to a $\sim -7\%$ change~\cite{Anastasiou:2009kn} 
to the gluon-fusion cross section.
At NNLO, contributions due to the top quark
have been evaluated as a systematic expansion around the infinite
top-quark mass limit~\cite{Pak:2009dg,Harlander:2009mq} finding
corrections of less than $1\%$. Light quark flavour effects, of which
the interference of diagrams with top and bottom quark loops is the most relevant,  
are not yet known at NNLO.  Assuming a typical NNLO K-factor as for
the top-quark contributions in the infinite top-quark limit, one can
expect a contribution of the order $\sim 7\%_{\rm NLO} \times
20\%_{K_{\rm NNLO/NLO}} \sim 1.5 \%$. This is similar to the $\pm 2\%$
scale-variation of the N$^3$LO corrections discussed above.  

A different class of contributions arise at two loops due to the quark coupling to electroweak (EW) vector bosons $V=W,Z$
through a quark loop, followed by the gauge coupling of the EW bosons to the Higgs boson.
The two-loop EW contributions due to light-quark flavours are dominant and were computed analytically in
Ref.~\cite{Aglietti:2004nj}. The two-loop amplitude was completed with
the addition of heavy-quark flavour contributions in Ref.~\cite{Actis:2008ug}. These two-loop EW
contributions are equal to  a $\sim +5.15\%$ increase of the leading order gluon-fusion cross section and a $\sim +2\%$ increase of the N$^3$LO
cross section. Electroweak production of a Higgs boson involving initial state quarks at the same power in all coupling constant was considered in ref.~\cite{Keung:2009bs} and was shown to be numerically small.

A pertinent question is then how large are the NLO QCD corrections to the two-loop EW contributions
due to light-quark flavours, given the experience from pure QCD corrections which are large at NLO and NNLO.
That would require combining the three-loop virtual corrections, recently computed~\cite{Bonetti:2016brm,Bonetti:2017ovy},
with the real emission contributions from two-loop four-point functions, which depend on four mass scales,
$s, t, m_H^2, m_V^2$, and are as yet unknown.

The NLO QCD corrections have been computed in the unphysical limit when the Higgs mass is much smaller
than the EW-boson mass $m_H \ll m_V$~\cite{Anastasiou:2008tj}, by means
of effective theory methods. In this limit, all mixed QCD-EW amplitudes 
factorise into a product of a Wilson coefficient and the same QCD
amplitude which emerges in pure QCD corrections in HEFT. 
Thus, in the $m_H \ll m_V$ limit, mixed QCD-EW corrections follow the same
perturbative pattern as in the pure QCD corrections and amount to 
$\sim +5\%$ of the all-orders QCD cross section.

There has been no progress towards a complete computation of the mixed
QCD-EW corrections for a long time until a recent breakthrough in
Ref.~\cite{Bonetti:2016brm,Bonetti:2017ovy}, which computes the three-loop
virtual corrections to that process for arbitrary values of the gauge
and Higgs boson masses.  Using these results, the same authors
computed with a full mass dependence the NLO QCD corrections 
in the soft gluon approximation~\cite{Bonetti:2018ukf}, for which the real emission contributions
are given by an almost universal factor. Ref~\cite{Bonetti:2018ukf}
could challenge earlier estimates of the mixed QCD-EW corrections
since the three-loop amplitude
contains a hard contribution which differs from the amplitude computed in the heavy mass
effective theory. The phenomenological outcome
of the work in  Ref.~\cite{Bonetti:2018ukf} was that the difference
due to the three-loop virtual contribution is small with respect to
the prescription of Ref.~\cite{Anastasiou:2008tj}.

However, the factorisation of mixed QCD-EW corrections can also
break down  due to hard real radiation for the physical values of the 
Higgs boson and the EW boson masses, $m_H > m_V$. 
While one should reasonably expect that this ``non-factorising'' 
contribution which is not captured by the soft approximation is not
large, the amount of the violation of the factorisation ansatz cannot be predicted by extrapolating from the 
un-physical region $m_H\ll m_V$, due to the presence of a threshold at $m_H = 2m_V$. 
It is therefore still necessary that the mixed  QCD-EW corrections be computed with a full mass dependence.  

A complete computation is challenging. 
So far, only the 	planar master integrals for the two-loop light-fermion EW corrections to Higgs plus jet production,
with arbitrary values of the gauge and Higgs boson masses, have been computed~\cite{Becchetti:2018xsk}.
In this paper, we consider the approximation of small EW boson masses, $m_V^2 \ll m_H^2, s, |t|$. This approximation shares more of the
complications of the full computation than the heavy mass $m_V \to
\infty$ limit.    For example, the number of loops is the same and a
non-factorisable part of the cross section from hard real radiation is
also present. We compute this small mass approximation of the cross section
analytically using the methods described in
Ref.~\cite{Dulat:2014mda}.  Then, we identify the parts of the cross section which are clearly
factorizable and compare it to the remainder.  In this respect, we point out that also
the second term in the threshold expansion (next-to-soft) also
factorises and we can therefore determine fully its dependence on the
mass of the electroweak gauge bosons.  

The scope of the calculation presented in this paper is triadic.
First, we provide a significant check of the
calculation of Ref.~\cite{Bonetti:2016brm,Bonetti:2017ovy} for  the
three-loop amplitude.  Second, our  analytic results offer a
check and/or a boundary condition for a future computation of the
cross section with arbitrary masses.  Third, we obtain for the first time a 
phenomenological estimate of the relative importance of the 
non-factorisable hard real radiation
contributions.

\section{Mixed QCD-EW corrections to the $gg \to H$ cross section}
\label{sec:xsection}

The hadronic cross section for Higgs boson production in the
gluon-gluon channel is
\begin{align}
	\sigma &= \int_0^1 \d x_1 \int_0^1 \d x_2\, f_{g/h_1}(x_1,\mu_F^2)\, f_{g/h_2}(x_2,\mu_F^2)\,
	 \hat\sigma_{gg}(S, m_H^2,
	 \mu_R^2,\mu_F^2)\,\nonumber\\
	 &= \tau\,\int_\tau^1 \frac{\d z}{z} \int_{\tau/z}^1 \frac{\d x_1}{x_1}\, f_{g/h_1}(x_1,\mu_F^2)\, f_{g/h_2}\left(\frac{\tau}{x_1 z},\mu_F^2\right)\,
	 \hat\sigma_{gg}(z,
	 \mu_R^2,\mu_F^2)\,, \label{eq:xsec}
\end{align}
where 
$m_H$ is the mass of the Higgs boson,
\begin{align}
 \tau= \frac{m_H^2}{S},\, \qquad  z=\frac{m_H^2}{s}=\frac{m_H^2}{x_1x_2S}=\frac{\tau}{x_1x_2}, 
\end{align}
$S$ is the squared hadron centre-of-mass energy,
$f_i(x)$ are the parton distribution functions and 
$\hat \sigma_{gg}$ the renormalised partonic cross section.

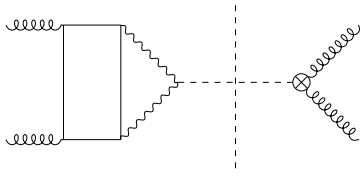
\begin{figure}[t!]
\begin{center}
\resizebox{0.31\textwidth}{!}{\begin{tikzpicture}
\begin{feynman}
\vertex (a1);
\vertex[draw,circle,fill=black,inner sep=0pt,minimum size=3pt,right=1cm of a1] (a2) ;
\vertex[draw,circle,fill=black,inner sep=0pt,minimum size=3pt,right=1cm of a2] (a3) ;

\vertex[below=2cm of a1] (b1);
\vertex[draw,circle,fill=black,inner sep=0pt,minimum size=3pt,below=2cm of a2] (b2) ;
\vertex[draw,circle,fill=black,inner sep=0pt,minimum size=3pt,below=2cm of a3] (b3) ;

\vertex[draw,circle,fill=black,inner sep=0pt,minimum size=3pt,below right = 1cm and 1cm of a3] (c1) ;
\vertex[draw,/tikzfeynman/crossed dot,right = 2cm of c1] (c2) {};

\vertex[below right = 1.0cm and 1.0cm of c2] (b4);
\vertex[above right = 1.0cm and 1.0cm of c2] (a4);
\vertex[below right = 1.5cm and 1cm of c1] (d1);
\vertex[above right = 1.5cm and 1cm of c1] (d2);
\diagram* {
	{[edges=gluon]
		(a1) -- (a2),
		(b1) -- (b2),
		(c2) -- {(a4),(b4)},		
	},
	{[edges=plain]
		(a2) -- (a3) -- (b3) -- (b2) -- (a2),
	},	
	{[edges=boson]
		(a3) -- (c1) -- (b3),
	},
	{[edges=scalar]
		(c1) --(c2),
		(d1) -- (d2),
	},
};
\end{feynman}
\end{tikzpicture}}
\end{center}
\caption{
\label{fig:one}
Representative Feynman diagram contributing to $\CG_\heft^{(0) \ast}\, \CG_{lf}^{(0)}$ of Eq.~(\ref{eq:heftew0}). The vertical dashed line 
represents the interference  between the
two-loop light quark form factor on the left,
and the top-loop form factor in HEFT on the right.}
\end{figure}

	\begin{figure}[t!]
	\begin{center}
		\begin{subfigure}[t]{0.31\textwidth}
			\resizebox{\textwidth}{!}{\begin{tikzpicture}
\begin{feynman}
\vertex (a1);
\vertex[draw,circle,fill=black,inner sep=0pt,minimum size=3pt,right=1cm of a1] (a2) ;
\vertex[draw,circle,fill=black,inner sep=0pt,minimum size=3pt,right=1cm of a2] (a3) ;
\vertex[draw,circle,fill=black,inner sep=0pt,minimum size=3pt,right=1cm of a3] (a4) ;

\vertex[below=2cm of a1] (b1);
\vertex[draw,circle,fill=black,inner sep=0pt,minimum size=3pt,below=2cm of a2] (b2) ;
\vertex[draw,circle,fill=black,inner sep=0pt,minimum size=3pt,below=2cm of a3] (b3) ;
\vertex[draw,circle,fill=black,inner sep=0pt,minimum size=3pt,below=2cm of a4] (b4) ;

\vertex[draw,circle,fill=black,inner sep=0pt,minimum size=3pt,below right = 1cm and 1cm of a4] (c1) ;
\vertex[draw,/tikzfeynman/crossed dot,right = 2cm of c1] (c2) {};

\vertex[below right = 1cm and 1cm of c2] (b5);
\vertex[above right = 1cm and 1cm of c2] (a5);

\vertex[below right = 1.5cm and 1cm of c1] (d1);
\vertex[above right = 1.5cm and 1cm of c1] (d2);

\diagram* {
	{[edges=gluon]
		(a1) -- (a2) -- (a3),
		(b1) -- (b2) -- (b3),
		(a2) -- (b2),
		(c2) -- {(a5),(b5)},
	},
	{[edges=plain]
		(a3) -- (a4) -- (b4) -- (b3) -- (a3),
	},	
	{[edges=boson]
		(a4) -- (c1) -- (b4),
	},
	{[edges=scalar]
		(c1) --(c2),
		(d1) -- (d2),
		},
};
\end{feynman}
\end{tikzpicture}}
			\caption{}
			\label{fig:2:subim1}
		\end{subfigure}
		\hspace{0.0225\textwidth}
		\begin{subfigure}[t]{0.31\textwidth}
			\resizebox{\textwidth}{!}{\begin{tikzpicture}
\begin{feynman}
\vertex (a1);
\vertex[draw,circle,fill=black,inner sep=0pt,minimum size=3pt,right=1cm of a1] (a2) ;
\vertex[draw,circle,fill=black,inner sep=0pt,minimum size=3pt,right=1cm of a2] (a3) ;

\vertex[below=2cm of a1] (b1);
\vertex[draw,circle,fill=black,inner sep=0pt,minimum size=3pt,below=2cm of a2] (b2) ;
\vertex[draw,circle,fill=black,inner sep=0pt,minimum size=3pt,below=2cm of a3] (b3) ;

\vertex[draw,circle,fill=black,inner sep=0pt,minimum size=3pt,below right = 1cm and 1cm of a3] (c1) ;
\vertex[draw,/tikzfeynman/crossed dot,right = 2cm of c1] (c2) {};

\vertex[draw,circle,fill=black,inner sep=0pt,minimum size=3pt,below right = 1cm and 1cm of c2] (b4) ;
\vertex[draw,circle,fill=black,inner sep=0pt,minimum size=3pt,above right = 1cm and 1cm of c2] (a4) ;
\vertex[below right = 1cm and 2cm of c2] (b5);
\vertex[above right = 1cm and 2cm of c2] (a5);

\vertex[below right = 1.5cm and 1cm of c1] (d1);
\vertex[above right = 1.5cm and 1cm of c1] (d2);
\diagram* {
	{[edges=gluon]
		(a1) -- (a2),
		(b1) -- (b2),
		(c2) -- {(a4),(b4)},
		(a5) --(a4) -- (b4) -- (b5),
	},
	{[edges=plain]
		(a2) -- (a3) -- (b3) -- (b2) -- (a2),
	},	
	{[edges=boson]
		(a3) -- (c1) -- (b3),
	},
	{[edges=scalar]
		(c1) --(c2),
		(d1) -- (d2),
	},
};
\end{feynman}
\end{tikzpicture}}		
			\caption{}
			\label{fig:2:subim2}
		\end{subfigure}
		\hspace{0.0225\textwidth} 
		\begin{subfigure}[t]{0.288\textwidth}
			\resizebox{\textwidth }{!}{\begin{tikzpicture}
\begin{feynman}
\vertex (a1);
\vertex[right=1cm of a1] (a2) ;
\vertex[right=0.5cm of a2] (a3) ;
\vertex[right=0.5cm of a3] (a4) ;

\vertex[below=2cm of a1] (b1);
\vertex[below=2cm of a2] (b2) ;
\vertex[below=2cm of a4] (b3) ;

\vertex[below right = 1cm and 1cm of a4] (c1) ;
\vertex[draw,/tikzfeynman/crossed dot,right = 2cm of c1] (c2) {};

\vertex[below right = 1.4cm and 1.4cm of c2] (b4);
\vertex[above right = 1cm and 1cm of c2] (a5) ;
\vertex[above right = 0.4cm and 0.4cm of a5] (a6);

\vertex[below right = 1.48cm and 1cm of c1] (d1);
\vertex[above right = 2cm and 1cm of c1] (d2);

\diagram* {
	{[edges=gluon]
		(a2) -- (a1),
		(b1) -- (b2),
		(b4) -- (c2) -- (a5) -- (a6),		
	},
	{[edges=plain]
		(a2) -- (a3) -- (a4) --(b3) -- (b2) -- (a2),
	},	
	{[edges=boson]
		(a4) -- (c1) -- (b3),
	},
	{[edges=scalar]
		(c1) --(c2),
		(d1) -- (d2),
	},
	(a5) -- [gluon,bend right] (a3),
};
\end{feynman}
%
%
%
%
%
%
%
%
\end{tikzpicture}}
			\caption{}
			\label{fig:2:subim3}
		\end{subfigure}
	\end{center}
	\caption{
		\label{fig:two}
		Examples of Feynman diagrams contributing to the $\ord(\as)$ corrections (\ref{eq:heftew1}) to the interference between the two-loop light-quark form factor and the top-loop form factor in HEFT. Fig.~\ref{fig:2:subim1}: Contribution to $ \CG_{\heft}^{(0) \ast}\, \CG_{lf}^{(1)}$.  Fig.~\ref{fig:2:subim2}: Contribution to $ \CG_{\heft}^{(1) \ast}\, \CG_{lf}^{(0)}$. Fig.~\ref{fig:2:subim3}: Contribution to $ \CG_{\heft-g}^{(0) \ast}\, \CG_{lf-g}^{(0)} $.
}
		\label{fig:diagrams_examples}
	\end{figure}

The NLO contributions to $\hat \sigma_{gg}$ due to mixed QCD-EW corrections are given in the heavy top mass limit,
up to $\mathcal O (\alpha_s^3 \, \alpha^2)$, by
\begin{align}
	\label{eq:general_decomposiction_xs}
	\hat\sigma_{gg}^{\textit{NLO,EW}} = \sigma_0\, \left[ \sigma_\heftew^{(0)} + \frac{\as}{\pi}\, \sigma_\heftew^{(1)} \right],\,
\end{align}
with
\begin{align}
	\sigma_\heftew^{(0)} &= 2\RE\left[ \CG_\heft^{(0) \ast}\, \CG_{lf}^{(0)}\right], \, \label{eq:heftew0}\\
	\sigma_\heftew^{(1)} &= 2\RE\left[ \CG_\heft^{(0) \ast}\, \CG_{lf}^{(1)} + \CG_\heft^{(1) \ast}\, \CG_{lf}^{(0)} 
	+ \CG_{\heft-g}^{(0) \ast}\, \CG_{lf-g}^{(0)} \right]. \label{eq:heftew1} 
\end{align}
Here $\CG_{lf}^{(0)}$ denotes the two-loop light-quark form factor and
$\CG_\heft^{(0)}$ is the top-loop form factor in the HEFT. Their interference is shown in Fig.~\ref{fig:one}.
$\sigma_\heftew^{(1)}$ are the NLO corrections to it,
as displayed in \fig{fig:two}. The overall normalization of the cross section \eqref{eq:general_decomposiction_xs}
\begin{align}
	\sigma_0 = \frac{\as^2}{576\pi v^2}
\end{align} 
is chosen such that the leading-order QCD contribution in HEFT reads
\begin{align}
	  \hat \sigma_{gg}^{LO,QCD}=\sigma_0|\CG_\heft^{(0)}|^2 =\sigma_0 \delta(1-z).
\end{align}
%
%
In order to study the different limits considered in this paper it is convenient to write the NLO mixed QCD-EW corrections by imposing the factorisation with respect to the leading-order contribution,
\begin{equation}
\label{eq:EW_NLO}
    \hat{\sigma}_{gg}^{\textit{NLO,EW}}=\sigma_0 \sigma_\heftew^{(0)} \left[\delta(1-z)+ \frac{\alpha_s}{\pi}\left(\eta_{gg}^{fact}+\delta(1-z)\eta_{gg}^{hard,V}+N_c\eta_{gg}^{hard,R}\right)\right],
\end{equation}
The leading order $\sigma_\heftew^{(0)}$ and the relative virtual corrections $\eta_{gg}^{hard,V}$ have been computed in \cite{Aglietti:2004nj} and \cite{Bonetti:2017ovy}  respectively, retaining the full dependence on the EW bosons masses $m_W$, $m_Z$, while the universal factorizable (see next section) contribution $\eta_{gg}^{fact}$ is given by,
\begin{align} 
	\eta_{gg}^{fact} =& 2N_c\, \bigg \{ \zeta_2 \delta(1-z) + 2 \left[ \frac{\log(1-z)}{(1-z)}\right]_+
	+ \left[ \frac{1}{1-z} \right]_{+}  \log \left( \frac{m_H^2}{\mu^2} \right)
	\nonumber \\ 
	+& \left( z (1-z) +\frac{(1-z)}{z} -1 \right)  \log \left( \frac{m_H^2(1-z)^2}{\mu^2} \right)+1
	\bigg \} + \frac{11}4\, \delta(1-z). \,\label{eq:ngg_fact} 
\end{align}
 The only contributions which are not known with full EW bosons mass dependence are the real emission corrections $\eta_{gg}^{hard,R}$.  
 
 The mixed QCD-EW corrections have been also estimated in the literature in Ref.~\cite{Anastasiou:2008tj} in the heavy EW boson-mass limit, and they correspond to
  \begin{align}
\label{eq:minftylimit}
 \lim_{m_W,m_Z\rightarrow \infty}\eta_{gg}^{hard,V}&=\frac{7}{6},\nonumber\\
 \lim_{m_W,m_Z\rightarrow \infty}\eta_{gg}^{hard,R}&=2\frac{ \left(z^2-z+1\right)^2 \log (z)}{(z-1) z}-\frac{11}{6}\frac{ (1-z)^3}{z}-2.
\end{align} 

 \section{The small-mass approximation}
\label{subsec:setup}

In this section, we consider the first term of a small EW-boson mass expansion, ${m_V^2\ll m_H^2, s, |t|}$.
That is equivalent to take the massless limit for the propagators of the gauge bosons $W$ and $Z$ running in the loop, 
while retaining their mass in the couplings to the Higgs boson.
More precisely, in the case of the $W$ boson,
we account for the first two generations of quarks, 
while for the $Z$ boson we include the bottom quark as well. 
We neglect diagrams involving the top quark. 
This is motivated by the fact that they contribute to the two-loop electroweak correction 
by just a few percent of the light-quark electroweak effects~\cite{Degrassi:2004mx} and we expect a similar pattern at one higher order in QCD perturbation theory.
All the light quarks are considered to be massless.
In addition, we consider only the contributions from the 
gluon-gluon initial state, which is dominant. 
Specifically, we consider the gauge invariant subset of diagrams with only up to one final state parton. 
Contributions with two final state partons are separately gauge invariant and will be considered in Ref.~\cite{Hirschi2018}.

We generate all contributing Feynman diagrams using QGRAF~\cite{Nogueira:1991ex} 
and perform the colour-, spinor- and tensor-algebra with standard methods.
Within the framework of reverse unitarity~\cite{Anastasiou:2002yz,Anastasiou:2002wq,Anastasiou:2003yy,Anastasiou:2003ds},
all resulting expressions can be directly simplified by using
integration by parts (IBP) identities~\cite{Chetyrkin:1981qh,Tkachov:1981wb}
and the reduction algorithm of Laporta~\cite{Laporta:2001dd}. 
The reduction is performed using KIRA~\cite{Maierhoefer:2017hyi} and FIRE5~\cite{Smirnov:2014hma}. 
The master integrals have been 
computed in Refs.~\cite{Chetyrkin:1981qh,Gorishnii:1989gt,Bekavac:2005xs,Heinrich:2007at,Heinrich:2009be,Lee:2010cga,Gehrmann:2010ue}.
and in Ref.~\cite{Dulat:2014mda} as part of the computation of the
inclusive Higgs cross-section at N$^3$LO in perturbative QCD.

The primary analytic result of this paper is the computation of the mixed QCD-EW corrections in the approximation of a small EW-boson mass (light boson), as defined above. We obtain, 

\begin{align}
\label{eq:EW_NLO_mless}
    \hat{\sigma}_{gg,lb}^{\textit{NLO,EW}}=&\sigma_0  f^{(0)}_{HTEW}(3\zeta_3-2) \delta(1-z)\nonumber\\
    + &\sigma_0 f^{(0)}_{HTEW}  \frac{\alpha_s}{\pi}\left[ (3\zeta_3-2) \eta_{gg}^{fact}+\delta(1-z)V_{gg,lb}^{hard}+N_c R_{gg,lb}^{hard}\right],
\end{align}
where,
\begin{align}
	f_{HTEW}^{(0)} = - 
	\frac{3\,\a^2\, v^2}{2\, \sin^4\theta_W\, m_H^2}\,
	\left[ \frac2{\cos^4\theta_W} \left( \frac{5}4 - \frac{7}3\, \sin^2\theta_W + \frac{22}9\, \sin^4\theta_W \right) + 4 \right], \,
\end{align}
and the overall factor $(3\zeta_3 - 2)$ is the small EW-boson mass approximation,
for both the $W$ and $Z$ bosons,
of the two-loop light-quark form factor, while the non-universal part of the virtual cross section is
\begin{align}
\label{eq:ngg_hardV}
	V_{gg,lb}^{hard} =& N_c \left(\frac{31 \zeta (3)}{4}-\frac{\pi ^2 \zeta (3)}{24}-\frac{15 \zeta (5)}{2}-\frac{17}{12}+\frac{\pi ^2}{3}\right) \nonumber \\ 
	+& \frac{1}{N_c} 
	\left( -\frac{19 \zeta (3)}{8}-\frac{\pi ^2 \zeta (3)}{24}+5 \zeta (5)-\frac{5}{3}+\frac{\pi ^2}{48}-\frac{\pi ^4}{1440}\right).\,
\end{align}
The real contributions, $R_{gg,lb}^{hard}$, can be expressed in terms of harmonic polylogarithms~\cite{Remiddi:1999ew}, defined iteratively as,
\begin{equation}
	H(a_1,\hdots,a_n,x)=\int_0^x dt f(a_1,t) H(a_2,\hdots ,a_n,t),
\end{equation}
with $a_i\in\{0,1,-1\}$ and
\begin{equation}
	f(0,x)=\frac{1}{x},\quad f(1,x)=\frac{1}{1-x},\quad f(-1,x)=\frac{1}{1+x}.
\end{equation}
If $a_i=0$ for all $i\in{1,\hdots,n}$ then,
\begin{equation}
	H(0_n,x)=\frac{\log(x)^n}{n!}.
\end{equation}
By using the following compact notation for the harmonic polylogarithms \cite{Remiddi:1999ew} of argument $1-z$,
\begin{equation}
H_{\hdots,\pm n,\hdots} \equiv H(\hdots,\underbrace{0,\hdots,0}_{\text{(n-1)-times}},\pm 1,\hdots,1-z),
\end{equation} 
we get, in the physical domain $z\in[0,1]$, the following real-valued expression,
\begin{align}
\label{eq:etaR}
R_{gg,lb}^{hard}&=
-3 (z-1) H_{2,3}+\frac{1}{2} (z-1) H_{3,2}+(z-1) H_{4,0}+(z-1) H_{4,1} \nonumber \\
&+(4-3 z) H_{1,1,3}+\frac{1}{2} (-7 z-6) H_{1,2,2}+(5 z+1) H_{1,3,0}\nonumber \\
&+(5 z+1) H_{1,3,1}+H_{2,1,2}-2 (z-1) H_{2,2,0}+(1-2 z) H_{2,2,1}\nonumber \\
&+\frac{1}{2} (z-1) H_{3,0,0}+(-4 z-1) H_{1,1,1,2}+\frac{1}{2} (4 z+11) H_{1,1,2,0}\nonumber \\
&+\frac{1}{2} (4 z+9) H_{1,1,2,1}-\frac{7}{2} (z+1)
H_{1,2,0,0}+\frac{1}{2} H_{1,2,1,1}\nonumber \\
&-3 (z-1) H_{2,0,0,0} -\frac{3}{2} (2 z-3) H_{1,1,0,0,0}+\frac{1}{2} (-8 z-3) H_{1,1,1,0,0} \nonumber \\
&+\frac{1}{2} \left(-z^2-7 z+\frac{2}{z}+7\right) H_{2,2}+\frac{1}{2} \left(z^2+6 z-\frac{2}{z}-3\right) H_{3,0} \nonumber \\
&+\frac{1}{2} \left(z^2+6 z-\frac{2}{z}-3\right) H_{3,1}+\frac{1}{6} \left(-3 z^2-21
z+4\right) H_{1,1,2} \nonumber \\
&+\frac{1}{4} \left(2 z^2+14 z-27\right) H_{1,2,0}+\frac{1}{12} \left(6 z^2+54 z-23\right) H_{1,2,1} \nonumber \\
&+\frac{1}{2} \left(-z^2-8 z+\frac{2}{z}+8\right) H_{2,0,0}+\frac{1}{4} \left(-2 z^2-16 z+7\right) H_{1,1,0,0} \nonumber \\
&+\frac{1}{2} (2 z-3) H_{1,3}+\frac{1}{2} (z-1) H_{2,1,1}-\frac{21}{4} H_{1,0,0,0}-z H_{1,1,1,0}  \nonumber \\
&+\left(\frac{1}{24} \left(9 z^2-118 z-68\right)+(7 z+8) \zeta_2\right) H_{1,2}\nonumber +\left(\frac{1}{12} \bigg(3 z^2-131 z  \right. \nonumber \\
& \left. +\frac{36}{z}+116\bigg)+9 (z-1) \zeta_2\right) H_{2,0}+\bigg(\frac{1}{24} \left(15 z^2-146 z-22\right)  \nonumber \\
&-2 (2 z+1) \zeta_2\bigg) H_{1,0,0}+\left(\frac{1}{2} \left(z^2-22 z\right)+(5 z-16) \zeta_2\right) H_{1,1,0}\nonumber \\
&+\left(\frac{1}{8} (z-1) \left(z+\frac{16}{z}-9\right)-(z-1) \zeta_2\right) H_{3}+\frac{1}{12} \bigg(-55 z+\frac{12}{z}\nonumber \\
&+67\bigg) H_{2,1}+\bigg(\left(z^2+9 z-\frac{2}{z}-9\right) \zeta_2+\frac{1}{24} (z-1) \bigg(20 z^2-3 z\nonumber \\
&-\frac{24}{z}-194\bigg)+2 (z+2) \zeta_3\bigg) H_{2}+\bigg(\frac{1}{2} \left(-z^2-4 z+44\right) \zeta_2\nonumber \\
&+\frac{20
	z^4-43 z^3-135 z^2-58 z+192}{24 (z-1)}+6 (2 z+1) \zeta_3\bigg) H_{1,0}-\frac{1}{48} z \log ^4(z) \nonumber \\
&+\left(\frac{1}{24} \left(-z^2+39 z+8\right)+\frac{1}{6} (-8 z-3) \zeta_2\right) \log ^3(z)+\frac{1}{16} (z-19)\nonumber \\
&\times (z-1) \log ^3(1-z) + \bigg(\frac{1}{48} (z-1) \left(20 z^2-3 z-\frac{24}{z}-32\right)\nonumber \\
&+\frac{1}{4} \left(-z-\frac{2}{z}-7\right) (z-1) \zeta_2\bigg) \log ^2(1-z)+\bigg(\frac{1}{4} \left(2 z^2+20 z-7\right) \zeta_2\nonumber \\
&+\frac{20 z^4-43
	z^3-275 z^2+222 z+52}{48 (z-1)}+(7 z+4) \zeta_3\bigg) \log ^2(z)\nonumber \\
& + \bigg(\frac{3 \left(z^3+5 z^2-11 z+\frac{2}{z}+3\right) \zeta_3}{2 (z-1)}+\frac{-14 z^3-115 z^2+212 z-\frac{48}{z}-35}{12 (z-1)}\nonumber \\
&-\frac{3}{2} (z-1) \left(z-\frac{2}{z}-22\right) \zeta_2\bigg) \log
(1-z)+\bigg(\frac{1}{4} \left(9 z^2-104 z+10\right) \zeta_2\nonumber \\
&+\frac{3 \left(3 z^3-11 z^2+31 z-19\right) \zeta_3}{2 (z-1)}+\frac{-17 z^3+87 z^2-135 z+47}{6 (z-1)}\nonumber \\
&-3 (2 z+1) \zeta_4\bigg) \log (z)+\frac{3}{4} \left(z^2+6 z-\frac{2}{z}-5\right) \zeta_4+\bigg(\frac{1}{4} (z-1) \nonumber \\
&\times\left(23 z-\frac{36}{z}-147\right)-6\bigg) \zeta_3+\frac{1}{8} (z-1) \left(-20 z^2+3 z+\frac{24}{z}-49\right) \zeta_2\nonumber \\
&+\frac{1}{12} (213-44 z) (z-1)+4 \ .
\end{align}

It is interesting to study the limiting behaviour of the real contributions to the cross section,
$R^{hard}_{gg}$, around $z=0$ and $z=1$.
More precisely, in the proximity of $z=1$
we find
\begin{align} \label{eq:etaRz1}
R_{gg,lb}^{hard}=&
(1-z) \left\{ 4 \zeta (3) +\frac{\pi ^4}{60} -\frac{17}{4}  +\left(3 \zeta (3) +5 -\frac{3 \pi ^2}{2}\right) \log (1-z)\right.\nonumber\\
 -&\left(\frac{\pi ^2}{12}-\frac{1}{2}\right) \log ^2(1-z) +\left. \frac{1}{4} \log ^3(1-z)\right\}+\mathcal{O}\left((1-z)^2\right),
\end{align}
which, together with the limit $z\to 1$ of \eq{eq:ngg_fact}, yields the next-to-next-to-leading-power corrections to the threshold limit, $z\to 1$,
while in the proximity of $z=0$,
\begin{align} \label{eq:etaRz0}
 R_{gg,lb}^{hard}=& \frac{1}{z}\left(2 \zeta (3)-\frac{\pi ^2}{3}\right)+\mathcal{O}(1),\,
\end{align}
which, together with the limit $z\to 0$ of \eq{eq:ngg_fact}, provides the high energy limit of the NLO corrections  
to the interference (\ref{eq:heftew0}). 

Finally we note that our small mass approximation is recovered from the general formula Eq.~(\ref{eq:EW_NLO}) by using the results of Ref.~\cite{Bonetti:2016brm,Bonetti:2017ovy}, where we take the limits,
\begin{align}
\label{eq:m0limit}
\lim_{m_W,m_Z\rightarrow 0}  \sigma_\heftew^{(0)}&= f^{(0)}_{HTEW} (3\zeta_3-2),\nonumber\\
 \lim_{m_W,m_Z\rightarrow 0} \eta_{gg}^{hard,V}&=V_{gg,lb}^{hard}/(3\zeta_3-2),
\end{align} 
and by identifying,
\begin{equation}
\label{eq:m0limitR}
\lim_{m_W,m_Z\rightarrow 0}\eta_{gg}^{hard,R}=R_{gg,lb}^{hard}/(3\zeta_3-2).
\end{equation}

Eq.~(\ref{eq:m0limit}) agrees with the results of
Ref.~\cite{Bonetti:2018ukf} when the latter are computed in the
small mass approximation. This is a significant check of the computations in 
Refs.~\cite{Bonetti:2017ovy,Bonetti:2016brm}. 
The behaviour of the factorizable soft and next-to-soft contributions
of Eq.~(\ref{eq:ngg_fact}) are in agreement with the expectations 
from the next-to-leading-power corrections~\cite{Low:1958sn,Burnett:1967km,DelDuca:1990gz,Bonocore:2015esa} to colour-singlet production from
gluon-gluon fusion at NLO~\cite{DelDuca:2017twk}. 

\section{Phenomenological analysis}
\label{sec:num_res}

Currently, the mixed QCD-EW contributions in the Higgs
inclusive cross section are estimated by computing their relative size
with respect to the leading order EW contributions in the limit of
$m_V  \to \infty$. These mass values are unphysical. However, it is
reasonable to expect that the relative size of the corrections is not 
very sensitive to the masses of the EW bosons and that
they can be estimated by choosing a convenient value, albeit
unphysical.  In the $m_V \to \infty$ limit,  the mixed QCD-EW
corrections factorize in terms of the square of a Wilson
coefficient times a partonic cross section in an effective theory
where the Higgs boson couples at tree level to gluons. The same type of
factorization (with a different Wilson coefficient) holds for the pure 
QCD cross section  in the HEFT, i.e. in the limit of an infinite top-quark mass.
Therefore, in these mass limits, one finds very similar QCD K-factors 
for the top-quark and the electroweak contributions to the Higgs
cross section.

It is very important to check the validity of the phenomenological
predictions for the mixed QCD-EW corrections, which are made by means
of the factorization hypothesis. Recently, the authors of Ref.~\cite{Bonetti:2017ovy}
computed the three-loop QCD-EW corrections and calculated numerically
their contribution to the Higgs cross section in the
soft-virtual approximation~\cite{Bonetti:2018ukf}.
This calculation replaces with an exact/physical result 
the contribution to the cross section which was approximated earlier
by the NLO term of the Wilson coefficient.
Ref.~\cite{Bonetti:2018ukf} found that the  approximation based on the
Wilson coefficient is phenomenologically good.  While this observation
strengthens the credibility of the existing phenomenological
predictions, it is still an open question whether the pattern of
perturbative corrections can be significantly altered in different
ways.  Modifications on the structure of the perturbative corrections 
are theoretically anticipated when the cross section is  evaluated
away from the $m_V \to \infty$ limit.  Specifically, in the
$\eta_{gg}^{hard,R}$ and $\eta_{gg}^{hard,V}$ parts of the
cross section.   In this section, we compare the numerical impact of the $m_V \to \infty$ limit with the $m_V \to 0$ limit. This is the diametric reverse limit of the
one used in previous estimates and it can reveal a potential breakdown
of the phenomenological assumptions of Ref.~\cite{Anastasiou:2008tj}.

We use the NNLO PDF4LHC15 set~\cite{Butterworth:2015oua}, and take $m_H = 125$~GeV, 
$m_W = 80.398$~GeV, $m_Z = 91.88$~GeV, $\sin^2(\theta_W) = 0.2233 $, $\alpha =1/128$,  $G_F=1.16639\times 10^{-5} /\text{GeV}^2$ and a center of mass 
energy of 13~TeV. We evolve the strong coupling constant $\alpha_s$ to NNLO. 
In order to obtain a fast and reliable numerical implementation of the real corrections $R_{gg,lb}^{hard}$,
we perform two power series expansions around the points $z=0$ and $z=1$. We evaluate the series expansion around $z=1$ in the interval $z\in\left[\frac{1}{2},1\right]$  and truncate the series at $\mathcal{O}((1-z)^{50})$.
Similarly, we evaluate the expansion around $z=0$ in the interval $z\in\left[0,\frac{1}{2}\right)$ and truncate the series at $\mathcal{O}(z^{50})$. 
In this way we achieve for $R_{gg,lb}^{hard}$ a numerical precision
of at least $10^{-10}$ within the full physical interval $z\in[0,1]$.

\begin{figure}
\centering
\includegraphics[width=0.8\textwidth]{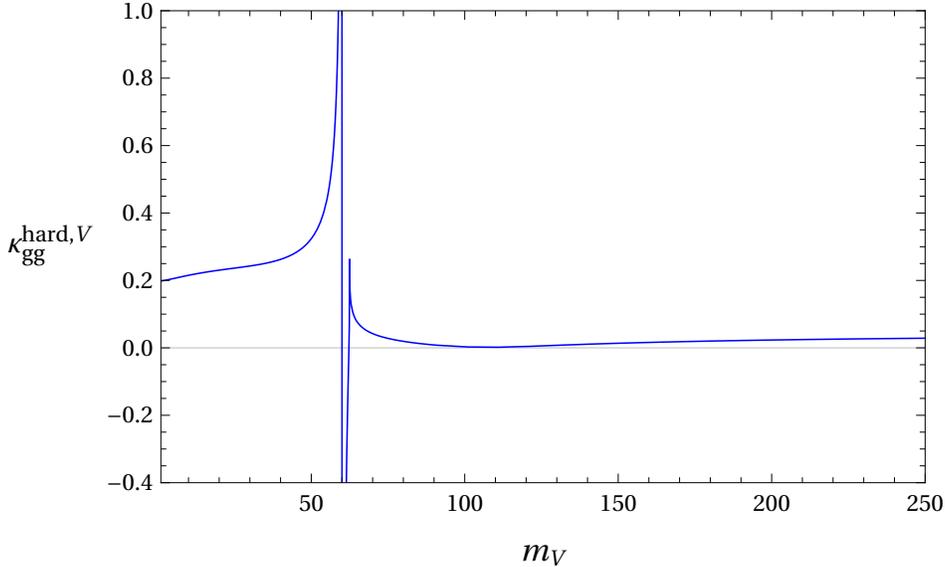}\\
\caption{Plot of $K_{gg}^{hard,V}$ for $m_V=m_W=m_Z$ and $\mu=m_H$}
\label{fig:etaVmV}
\end{figure}

\begin{table}
\begin{center}
\begin{tabular}{|c|c|c|c|c|c|c|}
\hline
 & & $\frac{\mu}{m_H}  = 2 $ & $\frac{\mu}{m_H}  = 1$ & $\frac{\mu}{m_H}  = \frac{1}{2}$ & $\frac{\mu}{m_H}  = \frac{1}{4}$
 		& $\frac{\mu}{m_H}  = \frac{1}{8}$ \\ \hline  
\multirow{2}{*}{$K_{gg}^{hard,R}$} & 
	$m_V\rightarrow 0$ & 0.14 & 0.15 & 0.18 & 0.21 & 0.25  \\ \cline{2-7}
	& $m_V\rightarrow \infty $ & 0.08 & 0.10 & 0.12 & 0.15 & 0.21 \\ 

\hline
\multirow{3}{*}{$K_{gg}^{hard,V}$} & 
	$m_V\rightarrow 0$ & 0.18 & 0.20 & 0.22 & 0.25 & 0.28 \\ \cline{2-7}
	& $m_V\rightarrow \infty $ & 0.04 & 0.04 &0.05 &0.05 &  0.06 \\ \cline{2-7}
	& phys &  0.014 & 0.015 &0.017  &0.019  & 0.022  \\ 
\hline
$K_{gg}^{fact}$ &  & 1.24 & 1.11 & 0.97 & 0.86 & 0.79  \\ 
\hline
\multirow{2}{*}{$K_{gg}^{\textit{NLO,EW}}$} & 
	$m_V\rightarrow 0$ & 1.56 & 1.46 & 1.37 & 1.31 & 1.33  \\ \cline{2-7}
	& $m_V\rightarrow \infty $ &  1.36 & 1.25 & 1.14 & 1.06 &  1.06 \\ \hline
\end{tabular}
\caption{$K$-factors of the various components of the NLO QCD-EW corrections within different approximations.}
\label{tab:K-factors}
\end{center}
\end{table}
We then decompose the NLO QCD-EW corrections as,
\begin{equation}
	\delta \sigma_{gg}^{\textit{NLO,EW}} = \sigma_{gg}^{fact}+ \sigma_{gg}^{hard,V} + \sigma_{gg}^{hard,R} \;, 
\end{equation}
where the three terms on the right-hand side correspond to the hadronic
cross section contributions from the NLO terms in \eq{eq:EW_NLO}. 
In Table~\ref{tab:K-factors},  we show the ratios of these terms to
the  leading order electroweak corrections in various approximations as a function of a common
factorization and renormalization scale $\mu$. 
We notice that the universal factorised contributions are dominant in both the approximations. 
The hard virtual contributions are a factor of 5 larger in the small mass approximation with respect to the heavy boson-mass approximation, while the physical values are about a factor of 2.5 smaller than the heavy mass approximation. The hard real contributions, are about 50\% larger in the small-mass approximation when compared to the large-mass approximation. It is also interesting to study the contribution 
of the virtual corrections for intermediate values of the boson mass. 
 In Fig.~\ref{fig:etaVmV} we present the plot of $ K_{gg}^{hard,V}$
as a function of the boson mass $m_V$ for $\mu=m_H$. On the left side of the plot (small mass approximation) the values approach $\frac{\alpha_s}{\pi} \frac{V_{gg,lb}^{hard}}{3 \zeta_3-2}=0.20$, while on the right side (heavy boson-mass approximation) the values approach $\frac{\alpha_s}{\pi} \frac{7}{6}=0.04$, which correspond to the values reported in Tab.~\ref{tab:K-factors}. 

Our results can be summarised by providing the values of the hadronic cross section in the different limits we have analysed. We set the reference scale to $\mu=m_H/2$. The pure QCD cross section in the HEFT is
\begin{equation}
\sigma_{gg}^{NLO,QCD}= 33.24 \text{ pb}.
\end{equation}
By including the mixed QCD-EW corrections Eq.~(\ref{eq:EW_NLO})  and taking $\eta^{hard,V/R}_{gg}$  in the heavy (hb) and light (lb) EW-boson limits as defined in Eqs.~(\ref{eq:minftylimit}, \ref{eq:m0limit}, \ref{eq:m0limitR}) we have, respectively 
\begin{align}
\sigma_{gg,hb}^{NLO,QCD-EW}&= 35.01 \text{ pb}, (= \sigma_{gg}^{NLO,QCD}+5.32 \%),\\
\sigma_{gg,lb}^{NLO,QCD-EW}&= 35.20 \text{ pb}, (= \sigma_{gg}^{NLO,QCD}+5.90 \%).
\end{align}
On the other hand, by taking the hard real contributions in the heavy and light EW bosons limits and keeping all the other contributions exact we have, respectively
\begin{align}
\sigma_{gg,R(hb)}^{NLO,QCD-EW}&= 34.98 \text{ pb}, (= \sigma_{gg}^{NLO,QCD}+5.23 \%),\\
\sigma_{gg,R(lb)}^{NLO,QCD-EW}&= 35.03 \text{ pb}, (= \sigma_{gg}^{NLO,QCD}+5.39 \%).
\end{align}
These results are in very good agreement with the (improved) soft-gluon approximation of Ref.~\cite{Bonetti:2018ukf}, where, by using our setup for PDFs and values of the parameters, the NLO QCD-EW corrections increase the QCD cross section by 5.33\% .

 With the recent calculation of the hard virtual contributions in Ref.~\cite{Bonetti:2017ovy,Bonetti:2018ukf}
with full $m_W,m_Z$ dependence, the only contribution which is known
in an unphysical limit is the hard real. This appears in our
calculation to be small. Moreover the full cross section appears to be relatively insensitive to different configurations of the boson masses for this contribution. We have therefore found no significant
deviations which invalidate the phenomenological assumptions 
made in the estimates of mixed QCD-EW 
corrections in Ref.~\cite{Anastasiou:2008tj}. Nevertheless, we would
like to note that this is not a ``bullet-proof''
exclusion of the possibility of larger hard real corrections for physical
EW-boson masses. A calculation for physical masses is therefore 
still motivated.

\section{Conclusions}
We have presented the calculation of the mixed QCD-EW corrections to the
Higgs production cross section in the gluon-gluon channel.
We neglect contributions from matrix elements with two final-state quarks.
These particular contributions are separately gauge invariant and are numerically sub-leading as will be shown in ref.~\cite{Hirschi2018}.
We work in the limit of massless propagators for the electroweak gauge bosons $W$ and $Z$.  
Besides providing a non-trivial check of the recent
exact calculation of the three-loop virtual corrections
\cite{Bonetti:2017ovy}, we could calculate, in the small-mass approximation, the
ratio of non-factorisable to factorisable contributions.  A large
ratio would challenge  the phenomenological assumptions which enter
the theoretical predictions for the inclusive Higgs boson production
cross section at the LHC.  We have reassuringly found a small
ratio.

Besides its phenomenological significance to test the validity of
prior theoretical predictions, our calculation will be a useful
stepping stone for a future full determination of the mixed QCD-EW
corrections for physical EW-boson masses.

\section*{Acknowledgements}
We would like to thank A.~Lazopoulos for useful discussions
on~\texttt{iHixs} and K.~Melnikov for clarifications on the results of
Ref.~\cite{Bonetti:2018ukf}.  This project has received funding from
the European Research Council (ERC) under grant agreement No 694712 (pertQCD),
the ETH Grant ETH-21 14-1 and the Swiss National Science Foundation (SNSF) under contracts 165772.
The research of BM was supported by the Pappalardo fellowship.
We thank the Galileo Galilei Institute for Theoretical Physics for the hospitality and the INFN for partial support during the completion of this work.


\bibliography{higgsrefs}


\end{document}